\documentclass[aps,reprint,showpacs]{revtex4-1}
\usepackage{bm}
\usepackage{amsmath}
\usepackage{amssymb}
\usepackage{graphicx}
\usepackage{color}
\usepackage{txfonts}
\usepackage[colorlinks=true,urlcolor=blue,citecolor=blue,linkcolor=blue,breaklinks=true]{hyperref}

\graphicspath{{fig/}}

\begin{document}

\title{$\eta$-pairing superfluid in periodically-driven fermionic Hubbard model with strong attraction}

\author{Sota Kitamura}
\affiliation{Department of Physics, University of Tokyo, Hongo, Tokyo 113-0033,
Japan}

\author{Hideo Aoki}
\affiliation{Department of Physics, University of Tokyo, Hongo, Tokyo 113-0033,
Japan}

\begin{abstract}
We propose a novel possibility of dynamically changing the pairing of superconductors from $s$ wave to $\eta$ pairing
(where the pairs condense at the Brillouin-zone corner momenta) by driving the system with ac fields. 
We consider a periodically-driven attractive Hubbard model in the strong-coupling regime,
and show that the pair-hopping and pair-repulsion terms in the effective Hamiltonian in the Floquet formalism are drastically renormalized in different manners between the two terms, 
which can change the ground states from an $s$-wave superconductivity to an $\eta$-pairing superconductivity or a charge-ordered phase. 
While in isolated systems such as cold atoms a simple quench scheme would not realize the dynamical phase transition into $\eta$ pairing, 
we show that there are pathways that realize the dynamical transition, 
where the field amplitude is varied via a charge-ordered phase as an intermediate state.
\end{abstract}

\date{\today}

\pacs{
74.20.Rp, % Pairing symmetries (superconductivity)
74.40.Gh, % Nonequilibrium processes; in superconductivity
67.85.-d  % Ultracold gases
71.10.Fd, % Lattice fermion models (Hubbard model, etc.)
}
%74.20.-z, % Hubbard model; superconductivity
%03.75.Kk, % Bose-Einstein condensation; dynamic properties
%74.20.Pq, % Electronic Structure; condensed matter; calculations in superconductivity
%74.20.Mn, % Fermi liquid; nonconventional mechanisms of superconductivity
%74.25.Gz, % Optical properties; of superconductors

\maketitle

\section{\label{sec:introduction}Introduction} 
Fascination with 
periodically-driven systems in condensed-matter and cold-atom physics is increasing as the become an important branch in nonequilibrium physics. 
Time-periodic external fields have turned out to make dramatic changes in physical properties
as captured by the Floquet formalism~\cite{Floquet1883}, which provide remarkable developments in both theoretical and experimental studies. 
A seminal example is the Floquet topological insulator~\cite{Oka2009,Kitagawa2011,Jotzu2014},
where a circularly polarized laser changes the ordinary system into 
a topological one.
Further exotic phenomena are proposed with the Floquet formalism~\cite{Grushin2014,Wang2014,Zheng2015,Nakagawa2015}, among which are 
a dynamical localization and accompanying photoinduced Mott transition~\cite{Dunlap1986,Eckardt2005,Tsuji2008,Mikami2015}, 
an attraction-repulsion conversion~\cite{Tsuji2011}, 
Floquet topological superconductivity~\cite{Kundu2013}, and 
a modulation of the exchange 
interaction~\cite{Mentink2014,Mikhaylovskiy2015}.
Periodic fields can be implemented 
by a laser light illuminated on electronic systems, or by shaking an optical lattice 
for cold-atom systems~\cite{Jotzu2014,Flaschner2016}.

Now, an intriguing question is, what would be the fate of a superconductor if we apply a periodic driving? 
One key interest is whether there exists a novel possibility of inducing an exotic superconductivity by driving a conventional superconductor.
In the present paper we provide an answer by revealing that an exotic, 
long-sought-after ``$\eta$ pairing"~\cite{Yang1989,Kantian2010,Rosch2008,Liberto2014,Bermudez2015} can indeed emerge in an attractive Hubbard model.
The $\eta$ pairing was originally proposed by Yang as an eigenstate possessing an off-diagonal long-range order in the Hubbard Hamiltonian~\cite{Yang1989}.
Behind this lies the fact that the Hubbard model has, in addition to the usual spin-SU(2) symmetry, 
another important symmetry called $\eta$-SU(2) with respect to a pseudospin $\hat{\bm{\eta}}$ [Eq.~(\ref{eq:eta-psin}) below].
The $\eta$ pairing is an exotic condensate with a nonzero expectation value of $\hat{\eta}^x+i\hat{\eta}^y$, which is a fingerprint of a Cooper pair with a 
nonzero momentum $\bm{Q}=(\pi,\pi,\cdots)$.
While the Fulde-Ferrell-Larkin-Ovchinnikov superconductivity~\cite{Casalbuoni2004,Bianchi2003,Lortz2007} is also characterized by a nonzero total momentum of a Cooper pair,
the $\eta$-pairing has a totally different mechanism as reflected in its ``maximized" momentum at the Brillouin-zone corners. Finite-momentum condensates have also been studied for bosonic systems~\cite{Kim2011,Soltan2012}.

\begin{figure}
\begin{centering}
\includegraphics[width=\hsize]{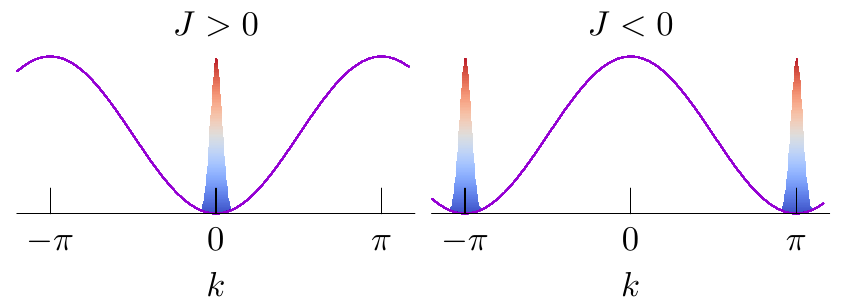}
\caption{\label{fig:bec}(color online) Schematic change of the ground state from $s$-wave pairing for a positive pair-hopping amplitude $J>0$ to $\eta$ pairing for a negative $J<0$ with the band bottom for doublons inverted.
The solid line represents the band dispersion, with the colored peak symbolizing the occupation of states, 
here illustrated in one dimension but conceivable in general dimensions. }
\end{centering}
\end{figure}

Here we consider the strong-coupling regime of the attractive Hubbard model, where fermions,
in equilibrium, form pairs (doublons) in real space and behave as an $s$-wave fermionic superfluid, as depicted in the left panel of Fig.~\ref{fig:bec}.
The possible candidates in real materials we have in mind~\footnote{
FeSe~\cite{Kasahara2014} would also be interesting in that this material is in the BCS-BEC crossover region with the size of the superconducting gap as large as the Fermi energy.}
include Ba$_{1-x}$K$_x$BiO$_3$ ~\cite{Meregalli1998} and doped fullerenes~\cite{Zhang1991}. 
More direct implementation of the model is expected to be cold-atom systems on optical lattices, where the interatomic interaction 
can be increased with the Feshbach resonance. Thus, hereafter, we refer to the superconductivity as the fermionic superfluid in a general context common to condensed-matter and cold-atom systems.

When we perform the strong-coupling expansion of the Hubbard model,
we have an effective Hamiltonian for doublons, which comprises pair-hopping and pair-repulsion terms.
In ordinary situations the pair-hopping amplitude $J$ is positive, 
for which the $s$-wave superconductivity is described as a condensation of doublons at the bottom of a bosonic band at $\bm{k}=\bm{0}$.
If one can invert the sign of the pair hopping, this would flip the band structure to give a new bottom at $\bm{k}=\bm{Q}$, 
where the $\eta$-pairing superconductivity is expected, as schematically depicted in Fig.~\ref{fig:bec}.
While one might imagine such a sign change would be unrealistic, this is in fact feasible in periodically-driven systems: 
We first show in Sec.~\ref{sec:low-energy-theory} that applying an ac laser field with a linear polarization (or shaking the optical lattice) changes the parameters drastically, 
which enables us to tune them in both magnitude and sign.

However, we have to be careful when discussing dynamical phase transitions
because the driven system should, in general, have a nonequilibrium energy distribution (such as a population inversion) 
and may not necessarily relax to the ground state.
This is important if we have cold atoms in an optical trap, where the system is isolated, so that a sudden flipping of
the band structure will induce an inverted population that is stable in the absence of dissipations. 

To overcome this, we propose to invoke a ``broken $\eta$-SU(2) symmetry," which will turn out to be a property peculiar to the attractive model (as opposed to the repulsive half-filled one). 
While the repulsive Hubbard model is mathematically equivalent to the attractive one at half filling in equilibrium, 
the equivalence is crucially broken once the driving field is turned on, 
which results in a breaking of the $\eta$-SU(2) symmetry. 
This brings about richer physics in the attractive case, 
where the pair hopping $J$ and pair repulsion $V$ are modified in different manners due to the phase in the hopping amplitude, 
which contrasts with the spin-SU(2)-invariant repulsive case~\cite{Mentink2014}.

Thanks to this, the dynamical instability arises, as we shall show in Sec.~\ref{sec:stability-analysis}: 
We derive a Gross-Pitaevskii-type equation of motion to obtain time evolutions and elementary excitations, 
which demonstrates that the superfluid phase is unstable in the regime where the pair repulsion is dominant, 
while the charge order becomes unstable when the pair hopping dominates.
These instabilities enable us to manipulate the order parameters, and we actually propose two protocols, 
a two-step quench and an adiabatic ramping, to realize the $\eta$-pairing superfluid, as prescribed in Sec.~\ref{sec:numerical-time-evolution}.

\section{\label{sec:low-energy-theory}Effective low-energy theory in the strong-coupling limit}
\subsection{Attractive Hubbard model and\\ underlying $\eta$-SU(2) symmetry}\label{sec:strong-coupling-A}
We start from the attractive Hubbard model on a cubic lattice, 
\begin{equation}
\hat{H}=-\sum_{ij}^{\text{n.n.}}\sum_{\sigma}t_{ij}\hat{c}_{i\sigma}^{\dagger}\hat{c}_{j\sigma}-U\sum_{i}\hat{n}_{i\uparrow}\hat{n}_{i\downarrow},
\label{eq:attractive-hubbard}
\end{equation}
in standard notations, where we set the hopping $t_{ij}=1$ hereafter.
For large enough attraction $U>0$, the ground state tends to maximize the double occupancy, and fermions behave as doublons at any filling.
The low-energy Hamiltonian for doublons $\hat{H}_\text{D}$ 
is described by the virtual processes in terms of the strong-coupling expansion (perturbation from $t_{ij}=0$),
\begin{equation}
\hat{H}_{\text{D}}=-J_{0}\sum_{ij}^{\text{n.n.}}\hat{c}_{i\uparrow}^{\dagger}\hat{c}_{i\downarrow}^{\dagger}\hat{c}_{j\downarrow}\hat{c}_{j\uparrow}+V_{0}\sum_{ij}^{\text{n.n.}}\frac{\hat{n}_{i}}{2}\frac{\hat{n}_{j}}{2},
\label{eq:attractive-heisenberg}
\end{equation}
where $J_{0}=2/U$ is the pair-hopping amplitude, while $V_{0}=2/U$ is the nearest-neighbor pair repulsion, with $\hat{n}_i/2$ being the number of doublons on the $i$th site.
The ground state is the condensate of doublons $\prod_{i}(1+e^{i\theta}\hat{c}_{i\uparrow}^{\dagger}\hat{c}_{i\downarrow}^{\dagger})|0\rangle$
in the mean-field approximation. 
We note that at half filling the charge-ordered state $\prod_{i{\in}A}(\hat{c}_{i\uparrow}^{\dagger}\hat{c}_{i\downarrow}^{\dagger})|0\rangle$ is also a ground state degenerate with the condensate.

Let us first recall the underlying $\eta$-SU(2) symmetry in the Hubbard model: The attractive Hubbard model is known to be equivalent to the repulsive one since the Shiba transformation~\cite{Shiba1972,Emery1976},
\begin{equation}
\hat{c}_{j\uparrow}^{\dagger} \leftrightarrow (-1)^{j}\hat{c}_{j\uparrow},
\label{eq:shiba-transformation}
\end{equation}
keeps the form of Eq.~(\ref{eq:attractive-hubbard}) except for the inverted sign of $U$.
This implies that Eq.~(\ref{eq:attractive-heisenberg}) is equivalent to a Heisenberg model, which can be seen if we rewrite Eq.~(\ref{eq:attractive-heisenberg}), up to a constant, as
\begin{equation}
\hat{H}_{\text{D}}=J_0\sum_{ij}^{\text{n.n.}}(\hat{\eta}^x_{i}\hat{\eta}^x_{j}+\hat{\eta}^y_{i}\hat{\eta}^y_{j})+V_0\sum_{ij}^{\text{n.n.}}\hat{\eta}^z_{i}\hat{\eta}^z_{j},
\label{eq:eta-heisenberg}
\end{equation}
where 
\begin{equation}
\hat{\eta}_{j}^{x}+i\hat{\eta}_{j}^{y} = (-1)^{j}\hat{c}_{j\uparrow}\hat{c}_{j\downarrow},\,\hat{\eta}_{j}^{z} = \frac{1}{2}(1-\hat{n}_{j\uparrow}-\hat{n}_{j\downarrow})\label{eq:eta-psin}
\end{equation}
is the pseudospin (Lie-algebra) operator that is transformed to the spin operator, $\hat{\bm{\eta}}_j\leftrightarrow\hat{\bm{S}}_j$, under the Shiba transformation.
Namely, the Hubbard model has, on top of the spin-SU(2) symmetry for $\hat{\bm{S}}=\sum_j\hat{\bm{S}}_j$, 
another $\eta$-SU(2) symmetry for $\hat{\bm{\eta}}=\sum_j\hat{\bm{\eta}}_j$, both in the attractive and repulsive cases.

However, we have to note that the constraint for fixing the particle number breaks the equivalence between attractive and repulsive cases except exactly at half filling 
since the chemical potential can be rewritten as a pseudo-Zeeman field $\propto\hat{\eta}^z$ along the $z$ axis.
More importantly, in the present context, an 
ac driving introduces a specific type of $\eta$-SU(2) breaking even in half-filled systems, as we shall reveal below.

\subsection{Floquet formalism}

Let us now turn to a periodic driving of the system. For lattice systems,
the Peierls substitution describes the coupling of fermions to ac laser fields as
\begin{equation}
t_{ij}\rightarrow t_{ij}e^{i\bm{A}(t)\cdot\bm{R}_{ji}}=t_{ij}\sum_{m=-\infty}^{\infty}i^m\mathcal{J}_{m}(\bm{A}\cdot\bm{R}_{ji})e^{-im\omega{t}},
\label{eq:peierls-substitution}
\end{equation}
where $\bm{R}_{ji}=\bm{R}_{j}-\bm{R}_{i}$, with $\bm{R}_{i}$ being the position of the $i$th site,
and $\bm{A}(t)=\bm{A}\cos\omega{t}$ is the vector potential of a linearly polarized laser.
On the right-hand side we have performed a Fourier transform,
where $\mathcal{J}_{m}$ is the $m$th Bessel function.
For cold-atom systems we shake optical lattices, where the vector potential is emulated by 
$\bm{A}(t)=\partial_{t}\bm{L}(t)$, with $\bm{L}(t)$ being the periodic motion of the lattice.

If we now apply the transformation~(\ref{eq:shiba-transformation}) to the present case with Eq.~(\ref{eq:peierls-substitution}), we have
\begin{equation}
e^{i\bm{A}(t)\cdot\bm{R}_{ji}}\hat{c}_{i\uparrow}^{\dagger}\hat{c}_{j\uparrow}\rightarrow-e^{i\bm{A}(t)\cdot\bm{R}_{ji}}\hat{c}_{i\uparrow}\hat{c}_{j\uparrow}^{\dagger}= e^{-i\bm{A}(t)\cdot\bm{R}_{ij}}\hat{c}_{j\uparrow}^{\dagger}\hat{c}_{i\uparrow};
\end{equation}
that is, the resultant Hamiltonian acquires a spin-dependent hopping as
\begin{equation}
-\sum_{ij}^{\text{n.n.}}t_{ij}e^{i\bm{A}(t)\cdot\bm{R}_{ji}}\hat{c}_{i\sigma}^{\dagger}\hat{c}_{j\sigma}\rightarrow
-\sum_{ij}^{\text{n.n.}}t_{ij}e^{-i\sigma\bm{A}(t)\cdot\bm{R}_{ji}}\hat{c}_{i\sigma}^{\dagger}\hat{c}_{j\sigma}.
\end{equation}
This degrades the equivalence between the repulsive and attractive cases and implies that the external field breaks the $\eta$-SU(2) symmetry.
The breaking of $\eta$-SU(2) becomes significant, especially in the strong-coupling limit,
since the $\eta$ spin becomes the relevant degree of freedom,
in terms of which the low-energy Hamiltonian~(\ref{eq:eta-heisenberg}) is expressed.
However, under a time-dependent drive, the energy is not conserved,
so that the usual perturbation scheme to derive Eq.~(\ref{eq:attractive-heisenberg}) is not applicable.

A key observation then is that we still have a discrete temporal translational symmetry 
since the ac laser field with a driving frequency $\omega$ is time periodic with a period $T=2\pi/\omega$.
Namely, we can employ Floquet's theorem~\cite{Floquet1883}:
While we always consider eigenstates of a generator $e^{-i\hat{H}t}$ for temporally translationally invariant systems,
here we can consider eigenstates of the generator for the discrete translation, $e^{-i\hat{F}T}=\mathcal{T}\exp[-i\int_0^Tdt\hat{H}(t)]$,
where $\mathcal{T}\exp[\bullet]$ is the time-ordered exponential. 
The eigenvalue of $\hat{F}$ is called the quasienergy $\epsilon$,
which is the temporal analog of the crystal momentum in Bloch's theorem 
and is defined over $(-\omega/2,\omega/2]$.
Then the quasienergy eigenstate is expressed as $\Psi(t)=\phi(t)e^{-i\epsilon{t}}$, with $\phi(t)=\phi(t+T)$.

By expanding $\hat{H}(t)$ and $\phi(t)$ in a Fourier series, we can convert the time-dependent Schr\"odinger equation into a time-independent one,
\begin{equation}
\sum_{l=-\infty}^{\infty}(\hat{H}_{m-l}-m\omega\delta_{m-l})\;\phi_{l}=\epsilon\phi_{m},\label{eq:extended}
\end{equation}
in the Floquet formalism for an extended Hilbert space,
where $\hat{H}_{m}$ and $\phi_{m}$ are the $m$th Fourier components of $\hat{H}(t)$ and $\phi(t)$, respectively.
Eigenvalues of this matrix are equivalent to those of $\hat{F}$ modulo $\omega$.

\subsection{Strong-coupling expansion in the Floquet formalism}

\begin{figure}
\begin{centering}
\includegraphics[width=\hsize]{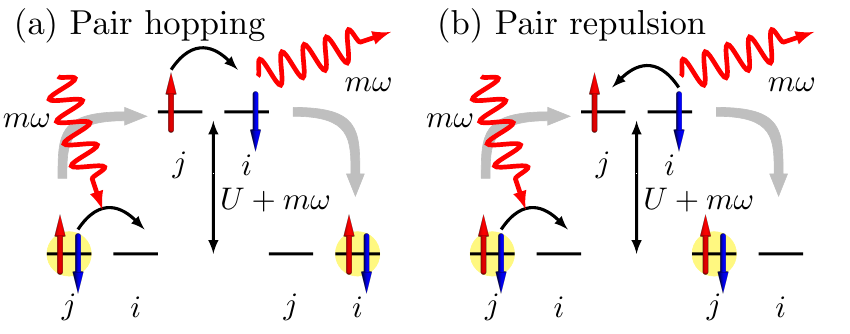}
\caption{\label{fig:virtual}(color online) Virtual processes for the (a) pair hopping and the (b) pair repulsion associated with photoabsorption/emission. 
Thick gray arrows represent virtual transitions, thin arrows represent fermion hopping, and wavy lines represent the coupling to photons.}
\end{centering}
\end{figure}

To explore the property of the Floquet equation (\ref{eq:extended}) for the periodically-driven Hubbard model,
we can perform the strong-coupling expansion in the extended Hilbert space,
where the hopping amplitude is decomposed into Fourier components $t_{ij}^{(m)}=i^m\mathcal{J}_{m}(\bm{A}\cdot\bm{R}_{ji})$ that 
contribute to virtual processes with an energy denominator $U+m\omega$, 
as depicted in Fig.~\ref{fig:virtual}.
These contributions via virtual Floquet states modify the original pair-hopping amplitude $J_{0}$ and the pair repulsion $V_{0}$ into
\begin{equation}
J_{\text{eff}}=\sum_{m=-\infty}^{\infty}(-1)^{m}\frac{2\mathcal{J}_{m}(A)^{2}}{U+m\omega},\, 
V_{\text{eff}}=\sum_{m=-\infty}^{\infty}\frac{2\mathcal{J}_{m}(A)^{2}}{U+m\omega}\label{eq:static-JV}
\end{equation}
in the time average. Here we take $\bm{A}=A(1,1,1)$. 

We can immediately notice that $J_{\text{eff}}$ and $V_{\text{eff}}$ are modified in different manners,
which is the very manifestation of the broken $\eta$-SU(2) symmetry due to the external field.
The difference comes from the interference of phase factors:
As depicted in Fig.~\ref{fig:virtual}, the pair-repulsion term is composed of a virtual fermion hopping from $j$ to $i$ and then back from $i$ to $j$,
which results in a cancellation of phases in the factor $|t^{(m)}_{ij}|^2$.
By contrast, the pair-hopping term is composed of a fermion hopping $j\rightarrow i$ followed by another $j\rightarrow i$, 
in which the resultant factor, $(t^{(m)}_{ij})^2$, acquires a phase.
Thus the two terms have distinct forms due to the phase factors.

This is quite unlike the periodically-driven repulsive Hubbard model at half filling~\cite{Mentink2014}, where the effective static Hamiltonian,
$\hat{F}_\text{rep}=V_\text{eff}\sum_{ij}^{\text{n.n.}}\hat{\bm{S}}_i\cdot\hat{\bm{S}}_j,$ has an isotropically modified exchange interaction $V_\text{eff}$.
In that case all the virtual processes in the strong-coupling expansion have the phase cancellation, leaving the spin-SU(2) symmetry intact.

\begin{figure*}
\begin{centering}
\includegraphics[width=\hsize]{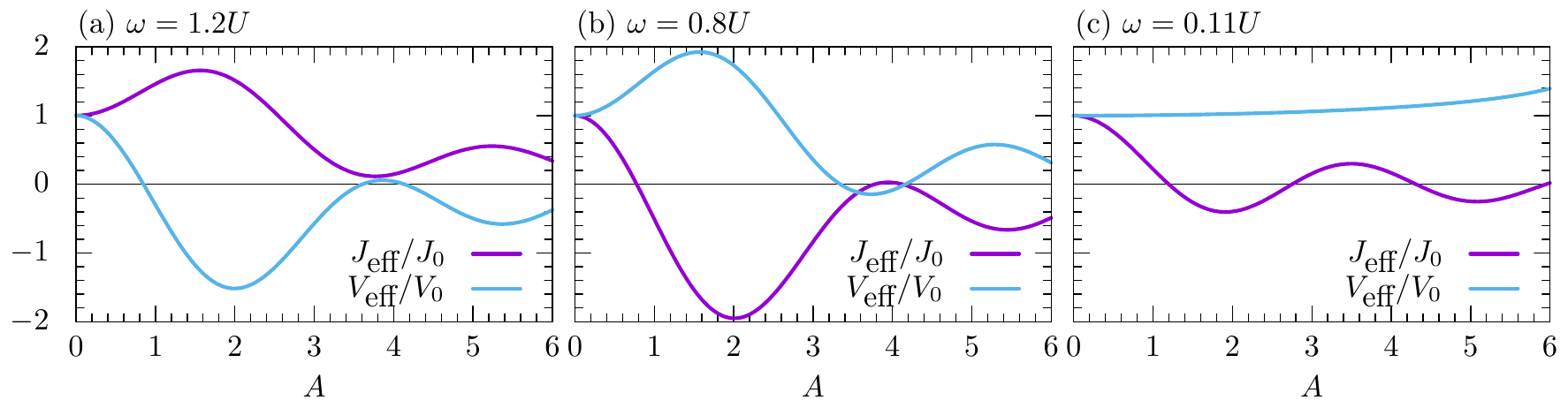}
\caption{\label{fig:params}(color online) Effective pair hopping $J_{\text{eff}}$ and pair repulsion $V_{\text{eff}}$ 
in the effective Floquet Hamiltonian against the amplitude of the drive $A$ 
for  (a) $\omega=1.2U$, (b) $\omega=0.8U$, and (c) $\omega=0.11U$.}
\end{centering}
\end{figure*}

Let us first examine the analytic behavior of the pair-hopping and pair-repulsion terms [Eq.~(\ref{eq:static-JV})] against the amplitude $A$ and frequency $\omega$ of the periodic drive.
For $\omega\gg{U}$, only the $m=0$ term survives, so that $J_{\text{eff}}\sim{V}_{\text{eff}}\sim(2/U)\mathcal{J}_0(A)^{2}$,
which just rescales the energy and nothing interesting happens.
A significant difference between $J_{\text{eff}}$ and $V_{\text{eff}}$ first occurs around $\omega\sim{U}$, with $\omega-U\gg1$, where we have asymptotically 
\begin{equation}
J_{\text{eff}} \sim -\frac{2\mathcal{J}_{1}(A)^{2}}{U-\omega},\,
V_{\text{eff}} \sim +\frac{2\mathcal{J}_{1}(A)^{2}}{U-\omega},
\end{equation}
except in the vicinity of zeros of $\mathcal{J}_{1}(A)$. 
If we now look at a numerical result for $J_{\text{eff}}$ and $V_{\text{eff}}$ against $A$ in Figs.~\ref{fig:params}(a) and \ref{fig:params}(b) 
for two cases of $\omega=1.2U$ and $\omega=0.8U$, 
the parameters indeed change in a manner dramatically sensitive to the value of $\omega$ 
due to the small denominator. 
For weak driving we can enhance $J_\text{eff}$ or $V_\text{eff}$, while for stronger drives we can even invert the signs of $J_{\text{eff}}$ and $V_{\text{eff}}$
and vary the ratio $J_{\text{eff}}/V_{\text{eff}}$ almost arbitrarily.

When we decrease $\omega$ even further, drastic changes in the behavior of the parameters occur in a large $A$ region when we are close to $\omega=U/m$,
which is due to the diverging contribution from $\mathcal{J}_{m}(A)^{2}/(U-m\omega)$.
For smaller values of $A$, we have $J_{\text{eff}}\sim(2/U)\mathcal{J}_{0}(2A)$ and $V_{\text{eff}}\sim2/U$ for $\omega/U\ll1$ in an $\omega/U$ expansion,
which is justified for contributions satisfying $m\omega/U\ll1$, but other contributions can be neglected due to $\mathcal{J}_{m}(A){\sim}A^{m}/(2^{m}m!)$ as far as the small-$A$ region is concerned.
This corresponds to the bosonic picture: 
When $\omega$ is small enough, doublons always behave as bosons, 
where the effective Hamiltonian looks like that of the shaken Bose-Hubbard model [with the hopping renormalized by $\mathcal{J}_0(2A)$].
This is shown in Fig.~\ref{fig:params}(c).
Thus the characteristic behavior for moderate $\omega/U$ regions found above comes from the fermionic nature of the doublon. 

\section{\label{sec:stability-analysis}Stability analysis of the\\ effective Hamiltonian}
We have revealed that the parameters in the effective Floquet Hamiltonian can be controlled, where we can even invert the sign of $J_\text{eff}$.
This urges us to consider an intriguing possibility of the exotic $\eta$-pairing superfluid~\footnote{For cold-atom systems,
the $\eta$-pairing ground state in the effective Floquet Hamiltonian is also investigated with different mechanisms~\cite{Liberto2014,Bermudez2015}.},
which can be expected from the emergence of a new bandbottom at $\bm{k}=\bm{Q}=(\pi,\pi,\pi)$ due to the band flip.

As stressed in the Introduction,
we have to take into account the nonequilibrium distributions  in discussing dynamical phase transitions,
which can make final states totally different from the equilibrium ones, especially for isolated systems as in trapped cold atoms. 
Thus we investigate in this section the stability of the $s$-wave superfluid under the effective Floquet Hamiltonian.
We shall see that the $s$-wave superfluid does not simply evolve to the $\eta$-pairing state even for the Hamiltonian with negative $J_\text{eff}$.
However, we shall then see that one can induce a collapse of the original superfluid when $V_\text{eff}$ is dominant,
which turns out to be a key for finally realizing the $\eta$-pairing state in Sec.~\ref{sec:numerical-time-evolution}.

\subsection{\label{sec:elementary-excitation}Elementary excitations}

In order to keep track of the fate of an $s$-wave superfluid,
here we consider its coarse-grained time evolution by the effective time-evolution operator $e^{-i\hat{F}t}$ with the effective Floquet Hamiltonian, 
\begin{equation}
\hat{F}=-J_\text{eff}\sum_{ij}^{\text{n.n.}}\hat{c}_{i\uparrow}^{\dagger}\hat{c}_{i\downarrow}^{\dagger}\hat{c}_{j\downarrow}\hat{c}_{j\uparrow}+V_\text{eff}\sum_{ij}^{\text{n.n.}}\frac{\hat{n}_{i}}{2}\frac{\hat{n}_{j}}{2}.\label{eq:effective-hamiltonian}
\end{equation} 
While the continuous-time evolution with this Hamiltonian differs from the actual one, they are identical at $t=nT$ since $e^{-i\hat{F}T}$ represents the actual time translation by definition.

Then we can discuss the time evolution in the mean-field approximation. Here we describe the superfluid state by a time-dependent ansatz, $|\Psi\rangle=\exp[\sum_{j}\Psi_{j}(t)\hat{c}_{j\uparrow}^{\dagger}\hat{c}_{j\downarrow}^{\dagger}]|0\rangle$, 
and take a variation 
$\delta\int{dt}\,\mathcal{S}(t)=0$, 
where $\mathcal{S}(t)=\langle\Psi|(\hat{F}-i\partial_{t})|\Psi\rangle/\langle\Psi|\Psi\rangle$.
This leads to an equation of motion, 
\begin{equation}
i\frac{\partial\Psi_{i}}{\partial t}=\sum_{j}^{\text{n.n.}}\frac{-J_\text{eff}\Psi_{j}+J_\text{eff}\Psi_{j}^{\ast}\Psi_{i}^{2}+2V_\text{eff}|\Psi_{j}|^{2}\Psi_{i}}{1+|\Psi_{j}|^{2}},\label{eq:Gross-Pitaevskii}
\end{equation}
which is analogous to the time-dependent Gross-Pitaevskii equation.  
We can easily confirm that the $s$-wave superfluid is a stationary solution of Eq.~(\ref{eq:Gross-Pitaevskii}) for arbitrary external fields,
given as $\Psi_0(t)=\Psi_0\exp\{-iz[(n-1)J_\text{eff}+nV_\text{eff}]t\}$, where $n$ is electron filling and $z=\sum_{j}^{\text{n.n.}}1$ is the coordination number.
This does not necessarily imply that the $s$-wave superfluid is always stable, as shown below.

The stability of the $s$-wave superfluid is examined by adding a small perturbation $\delta\Psi$ to see how it grows.
By denoting 
\begin{equation}
\Psi_i(t)=\Psi_0(t)\left[1+\sum_{\bm{k}}\delta\Psi_{\bm{k}}(t)e^{i\bm{k}\cdot\bm{R}_i}\right],
\end{equation}
Eq.~(\ref{eq:Gross-Pitaevskii}) is linearized into
\begin{equation}
i\frac{\partial}{\partial t}
\begin{pmatrix}
1 & 0\\
0 & -1
\end{pmatrix}
\begin{pmatrix}
\delta\Psi_{\bm{k}}\\
\delta\Psi_{-\bm{k}}^{\ast}
\end{pmatrix}
=
J_\text{eff}
\begin{pmatrix}
\mu_{\bm{k}}+\sigma_{\bm{k}} & \sigma_{\bm{k}}\\
\sigma_{\bm{k}} & \mu_{\bm{k}}+\sigma_{\bm{k}}
\end{pmatrix}
\begin{pmatrix}
\delta\Psi_{\bm{k}}\\
\delta\Psi_{-\bm{k}}^{\ast}
\end{pmatrix},
\label{eq:linearized-equation}
\end{equation}
where $\mu_{\bm{k}}=2\sum_{j}^{\text{n.n.}}\sin^{2}(\bm{k}\cdot\bm{R}_{ji}/2)$ and 
$\sigma_{\bm{k}}=n(1-n/2)(1+V_\text{eff}/J_\text{eff})\sum_{j}^{\text{n.n.}}\cos(\bm{k}\cdot\bm{R}_{ji})$.
The diagonalization of Eq.~(\ref{eq:linearized-equation}) yields a dispersion relation for elementary excitations,
with an energy 
\begin{equation}
\omega(\bm{k})=J_{\text{eff}}\sqrt{\mu_{\bm{k}}(\mu_{\bm{k}}+2\sigma_{\bm{k}})},
\label{eq:landau-instability}
\end{equation}
when $\mu_{\bm{k}}+2\sigma_{\bm{k}}>0$~\footnote{
There is another solution, $-\omega(\bm{k})$, but it has a negative Bogoliubov norm $|\delta\Psi_{\bm{k}}|^{2}-|\delta\Psi_{-\bm{k}}^{\ast}|^{2}$.
It is physically equivalent to the solution $\omega(\bm{k})$ with a positive Bogoliubov norm.}.

If $J_\text{eff}$ is changed to negative where the $\eta$-pairing ground state is expected, the excitation energy $\omega(\bm{k})$ becomes negative.
This will induce a Landau instability when the system is coupled to a heat bath.
In isolated systems, on the other hand, the $s$-wave superfluid will be retained under the evolution with the effective Hamiltonian.

Now we can note that $\mu_{\bm{k}}+2\sigma_{\bm{k}}$ is not necessarily positive:
Indeed, it becomes negative for, e.g., $\bm{k}=\bm{Q}$ when $V_\text{eff}/J_\text{eff}>1$ at half filling.
When $\mu_{\bm{k}}+2\sigma_{\bm{k}}<0$, the solution becomes 
\begin{equation}
\omega(\bm{k})=\pm iJ_{\text{eff}}\sqrt{\mu_{\bm{k}}|\mu_{\bm{k}}+2\sigma_{\bm{k}}|}.
\label{eq:dynamical-instability}
\end{equation}
The imaginary eigenvalue signals a dynamical instability; that is, a small deviation from an $s$-wave superfluid grows exponentially.
These instabilities, the Landau and dynamical ones, are analogous to those in Bose-Einstein condensates in an optical lattice~\cite{Wu2001}.

\subsection{\label{sec:classical-formulation}Hamiltonian mechanics formulation}

To grasp the physical origin and the fate of the dynamical instability,
we can consider a special case where the initial state is uniform on each of $A$ and $B$ sublattices in a bipartite lattice.
Then Eq.~(\ref{eq:Gross-Pitaevskii}) becomes coupled equations for $\Psi_A(t)$ and $\Psi_B(t)$.
We can then capture the quantities of physical interest as the differences in the doublon density $\rho_Q$ and in the phase $2\theta$ between the two sublattices,
\begin{align}
\rho_Q=&\frac{|\Psi_A(t)|^2}{1+|\Psi_A(t)|^2}-\frac{|\Psi_B(t)|^2}{1+|\Psi_B(t)|^2}, \\
2\theta=&\arg \Psi_A(t)-\arg \Psi_B(t).
\end{align}
The superconducting amplitude on each site $|\langle\hat{c}_{j\downarrow}\hat{c}_{j\uparrow}\rangle|$ is expressed as $\{1-[\rho_Q+(-1)^j(n-1)]^2\}^{1/2}/2$. 
With these variables, Eq.~(\ref{eq:Gross-Pitaevskii}) is simplified into
\begin{equation}
\frac{d\rho_{Q}}{dt}=\frac{\partial H}{\partial\theta},\,\;\frac{d\theta}{dt}=-\frac{\partial H}{\partial\rho_{Q}},
\label{eq:canonical}
\end{equation}
where
\begin{equation}
H=\frac{z}{2}(n^2-\rho_{Q}^{2})\left[V_{\text{eff}}-\sqrt{\frac{(2-n)^{2}-\rho_{Q}^{2}}{n^{2}-\rho_{Q}^{2}}}J_{\text{eff}}\cos2\theta\right].
\label{eq:classical-hamiltonian}
\end{equation}
Namely, the system obeys a Hamiltonian mechanics with $\rho_Q$ and $\theta$ acting as canonical variables for a ``classical" Hamiltonian $H$. 

\begin{figure*}
\begin{centering}
\includegraphics[width=\hsize]{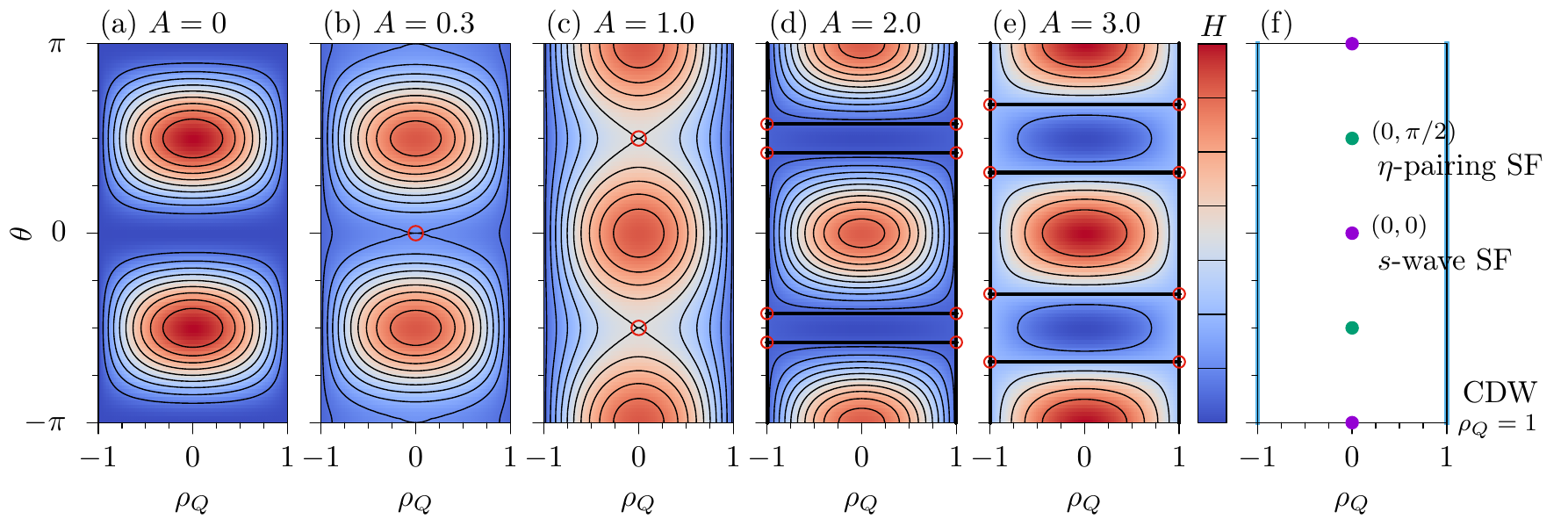}
\caption{\label{fig:contour}(color online) (a)-(e) Contours of 
$H\propto(1-\rho_{Q}^2)(V_\text{eff}-J_\text{eff}\cos2\theta)$ [Eq.~(\ref{eq:classical-hamiltonian})]
against $\rho_{Q}$ and $\theta$ for various values of $A$ with a fixed $\omega=0.8U$ at half filling.
Corresponding $(J_\text{eff},V_\text{eff})$ are shown in Fig.~\ref{fig:phase}.
Saddle points are marked by red circles, and the contours with $H=0$ are depicted with thick lines in (d) and (e).
(f) The $s$-wave superfluid corresponds to $(\rho_Q,\theta)=(0,0),(0,\pi)$, the $\eta$-pairing superfluid corresponds to $(\rho_Q,\theta)=(0,\pm\pi/2)$, and charge order corresponds to $\rho_Q=\pm1$.
}
\end{centering}
\end{figure*}

\begin{figure}
\begin{centering}
\includegraphics[width=\hsize]{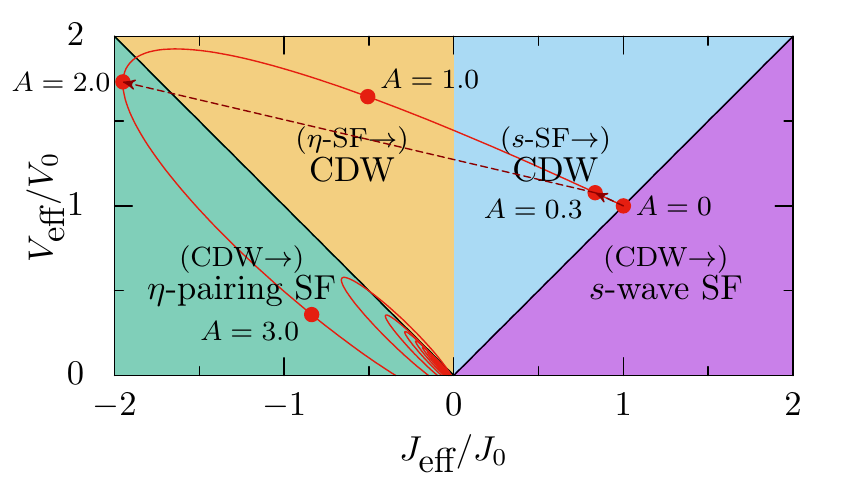}
\caption{\label{fig:phase}(color online) The phase diagram of the present system at half filling 
with phase boundaries depicted with solid black lines. 
Annotation above the name of a phase such as ``(CDW$\rightarrow$)" indicates that the dynamical instability into the phase occurs from a CDW phase.  The CDW phase comprises two regions (displayed with different colors) according to the dynamical instability they have. 
A red curve represents a locus of $(J_\text{eff}(A),V_\text{eff}(A))$ when $A$ is adiabatically varied for $\omega=0.8U$.
The dashed darkred arrow indicates the path in a quench protocol used in Fig.~\ref{fig:biquench}.}
\end{centering}
\end{figure}

Since Eq.~(\ref{eq:canonical}) has the form of a canonical equation of motion, $H$ is conserved under the time evolution;
that is, the system should evolve along a contour of $H$, which results in a periodic motion.
Figure~\ref{fig:contour} shows the contour map against $\rho_{Q}$ and $\theta$ for various values of $A$ with a fixed $\omega=0.8U$. 
We also display corresponding $(J_\text{eff}(A),V_\text{eff}(A))$ on the phase diagram against $J_\text{eff},V_\text{eff}$ in Fig.~\ref{fig:phase},
along with the trajectory when $A$ is varied.
Here we consider a half-filled system ($n=1$).

On the contour maps, the $s$-wave superfluid is represented by the origin, $(\rho_Q,\theta)=(0,0)$ [and $(0,\pi)$].
The $\eta$-pairing superfluid corresponds to the points $(\rho_Q,\theta)=(0,\pm\pi/2)$,
while the charge-ordered phases are delineated by $\rho_Q=\pm1$, as displayed in Fig.~\ref{fig:contour}(f).

In the absence of the external field [Fig.~\ref{fig:contour}(a)], $(\rho_Q,\theta)=(0,0)$ is the fixed point (bottom of the energy profile),
which indicates the stability of the $s$-wave superfluid in equilibrium. 
A parameter quench toward $J_\text{eff}<0$ changes the contour map to those shown in Figs.~\ref{fig:contour}(c)-\ref{fig:contour}(e).
There, the $s$-wave superfluid is retained as a fixed point (now at the top of the energy profile), so that a dynamical phase transition is not triggered, 
although the Landau instability would be induced if one attaches a heat bath [see Eq.~(\ref{eq:landau-instability})].
We can, however, notice that a quench toward a region with $V_{\text{eff}}/J_{\text{eff}}>1$ (blue region in Fig.~\ref{fig:phase}) 
turns the energy bottom into a saddle point as shown in Fig.~\ref{fig:contour}~(b).
This signifies an emergence of the dynamical instability with the system starting to evolve along the emergent contour that starts from the saddle point  [see Eq.~(\ref{eq:dynamical-instability})].

Namely, the dynamical instability of $s$-wave superfluid results in the collapse of the superfluid into an exponential growth of the charge-density wave (CDW) $\rho_Q$.
Because of the integrability of Eq.~(\ref{eq:classical-hamiltonian}), the motion is periodic;
the system eventually comes back to the initial $s$-wave superfluid state.
This feature disappears in nonintegrable cases, e.g., for an initial state with an inhomogeneity.
In such cases the system is expected to be in an equilibrium state for the effective Hamiltonian.
We note that this is an approximate description for an evolution over a finite time duration 
since in general the final state of the Floquet system after infinite time evolution should tend to be an infinite-temperature state
when the higher-order terms in the expansion are considered~\cite{DAlessio2014,Lazarides2014,Mori2016}.

Now, we note that various other saddle points exist besides that for the $s$-wave fluid, as marked with red circles in Fig.~\ref{fig:contour}.
These are crucial in manipulating the order parameters, as we shall see in the next section.
For instance, Fig.~\ref{fig:contour}(c) implies the existence of a dynamical instability from an $\eta$-pairing superfluid [$(\rho_Q,\theta)=(0,\pm\pi/2)$] toward a charge-density wave.
On the other hand, the saddle points in Figs.~\ref{fig:contour}(d) and \ref{fig:contour}(e) signify an instability from a charge-density wave ($|\rho_Q|=1$) into the desired $\eta$-pairing superfluid 
since $|\rho_Q|$ with $\theta\simeq\pm\pi/2$ decays along the emergent contour branching off the saddle point (horizontal thick lines) in Figs.~\ref{fig:contour}(d) and \ref{fig:contour}(e).

These instabilities are summarized in Fig.~\ref{fig:phase} as the phase diagram classified by the dynamical instabilities as well as the ground states.
There, various regions are characterized by the types of dynamical instabilities described above.
We also display the trajectory that represents $(J_\text{eff}(A),V_\text{eff}(A))$ when $A$ is adiabatically varied for $\omega=0.8U$.
We can see that the trajectory starts from the $s$-wave superfluid via the region for instabilities into CDW,
then plunges into another region for instabilities into the $\eta$ pairing. 

Incidentally, if one increases $A$ further, the trajectory whirls into $(J_\text{eff},V_\text{eff})=(0,0)$, as shown in Fig.~\ref{fig:phase}.
While $V_\text{eff}<0$ is not shown in the phase diagram, a phase separation is expected to occur, which cannot be described in the uniform formulation adopted here.

\section{\label{sec:numerical-time-evolution}Time-Evolution Protocols}
We have seen that, while a parameter region for the $\eta$-pairing ground state exists,
the $s$-wave superfluid does not evolve to the $\eta$-pairing superfluid. 
However, we find that a dynamical-instability channel toward the $\eta$ pairing (from CDW rather than from the $s$ wave) does exist,
along with a channel from the $s$-wave superfluid into the charge-ordered phase.

This opens a way, by a judicious choice of the temporal variation of the field amplitude,
to induce the evolution ending up with $\eta$ pairing via other phases.
The simplest way is to consider a two-step quench of the field amplitude, with
the first quench inducing a dynamical instability into a CDW state,
followed by another instability from CDW to $\eta$ pairing.
We describe this scheme in Sec.~\ref{sec:twostep}.

Alternatively, we can consider adiabatic ramping (as opposed to quench) of the field amplitude.
This approach gives a deeper insight in terms of the classical Hamiltonian mechanics introduced in the previous section 
because one can apply the adiabatic theorem for classical mechanics~\cite{Landau1977,Liu2002}.
We shall describe this scheme in Sec.~\ref{sec:adiabatic}.

Hereafter we concentrate on the half-filled case,
but we have confirmed that deviations from half filling do not significantly affect the qualitative behavior. 
Let us also comment on the time-dependent formalism with a nonperiodic modulation, which we adopt below.
Qualitatively, a temporal variation of the field amplitude $A(t)$ can be described by a temporal variation of the effective parameters $J_\text{eff}(A(t)),V_\text{eff}(A(t))$.
On the other hand, when the Hamiltonian is nonperiodic in time,
the Floquet theorem is no longer applicable, and the effective Hamiltonian cannot be interpreted as a generator of a discrete time translation. 
Still, we can take an approach with an effective Hamiltonian by considering the general time-dependent version of the strong-coupling expansion for an arbitrary vector potential. 
As we show in the \hyperref[sec:cano]{Appendix}, the system evolves with time-dependent parameters $J_{ij}(t),V_{ij}(t)$ [Eqs.~(\ref{eq:J-full}),~(\ref{eq:V-full})] 
in place of $J_\text{eff}(A(t)),V_\text{eff}(A(t))$.
In the following we adopt this formalism for numerical calculations.

\subsection{\label{sec:twostep}Two-step quench in the field amplitude}
Now let us devise a two-step quench protocol to obtain the $\eta$-pairing superfluid. We keep track of the evolution of the candidate order parameters, 
\begin{align}
\Delta_{0}&=\frac{2}{N}\sum_{j}\langle\hat{c}_{j\downarrow}\hat{c}_{j\uparrow}\rangle  &\text{($s$-wave pairing)}, \\
\Delta_{Q}&=\frac{2}{N}\sum_{j}(-1)^{j}\langle\hat{c}_{j\downarrow}\hat{c}_{j\uparrow}\rangle &\text{($\eta$ pairing)}, \\
\rho_{Q}&=\frac{1}{N}\sum_{j, \sigma}(-1)^{j}\langle\hat{c}_{j\sigma}^{\dagger}\hat{c}_{j\sigma}\rangle &\text{(charge-density wave)}.
\end{align}
We set $U=10$ and take an initial state with a tiny inhomogeneity as
$\Psi_i(0)=1+0.001\cos(\bm{q}\cdot\bm{R}_i)$, with $\bm{q}=(\pi/6,\pi/6,\pi/6)$.

\begin{figure}
\begin{centering}
\includegraphics[width=\hsize]{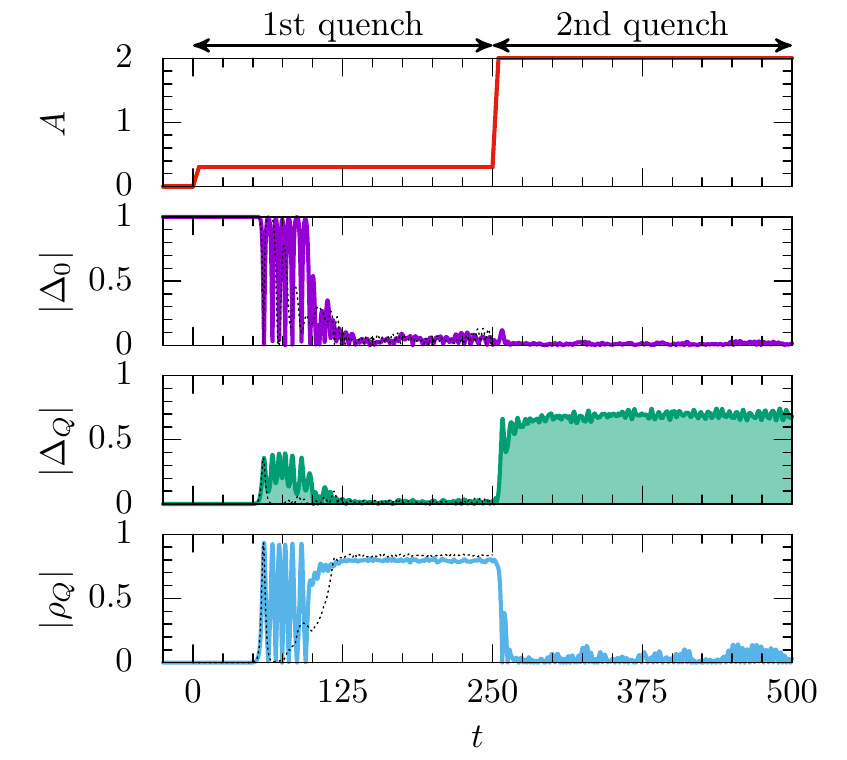}
\caption{\label{fig:biquench}(color online) Time evolution of the order parameters in a two-step ($t\lessgtr250$) quench protocol (see text) with $U=10$, $\omega=8=0.8U$.
The region where the $\eta$-pairing order parameter $\Delta_Q$ develops is highlighted in light green.
Black dotted lines for $t<250$ represent the time evolutions by the effective static Hamiltonian [Eq.~(\ref{eq:effective-hamiltonian})].
}
\end{centering}
\end{figure}

As the first quench, we turn on the ac field at $t=0$ with $\bm{A}(t)=A(t)(1,1,1)\sin\omega{t}$ with $A=0\rightarrow0.3$ and $\omega=8=0.8U$ as in Fig.~\ref{fig:phase} (first part of the dashed arrow).
In this situation the dynamical instability occurs from the $s$-wave superfluid into CDW,
as indicated by the blue region ($V_\text{eff}/J_\text{eff}>1$) in Fig.~\ref{fig:phase} [see also the discussion at the end of Sec.~\ref{sec:elementary-excitation}].
The numerical result for the time evolution is displayed for $t<250$ in Fig.~\ref{fig:biquench}.
A rapid growth of $\rho_{Q}$ from a small charge inhomogeneity included in the initial state is clearly seen at around $t\simeq50$, 
which relaxes to a steady charge order for $t\gtrsim125$ after a transient dynamics.

For comparison we also show the time evolution with the effective static Hamiltonian [Eq.~(\ref{eq:effective-hamiltonian})] for $A=0.3$. 
We can see that the collapse of the initial order and subsequent convergence to a final state are roughly described, although some deviations exist in the transient dynamics.

Then we perform the second quench (second part of the dashed arrow in Fig.~\ref{fig:phase}),
now from $A=0.3$ to $2.0$ into the region (green region in Fig.~\ref{fig:phase}),
where the charge order is expected to collapse toward the $\eta$-pairing superfluid.
The numerical result in Fig.~\ref{fig:biquench} for the time evolution after the quench at $t=250$ clearly shows that this is indeed the case,
with a sizable final amplitude close to unity for the $\eta$-pairing superfluid.
This strikingly contrasts with a simple quench (e.g., $A=0\rightarrow2.0$) in which no evolutions of the order parameters occur.

We can see that dominant order parameters evolve as 
(i) initially, $|\Delta_0|\simeq1$, then (ii) $|\rho_Q|\rightarrow0.8$ in the first quench,
and (iii) $|\Delta_Q|\rightarrow0.7$ in the second quench. 
The reason why $|\rho_Q|$ and $|\Delta_Q|$ do not reach unity 
is a heating effect, which increases for a larger change in $|V_\text{eff}/J_\text{eff}|$,
so that the ratio should be kept closer to unity in the quench protocol if we want to obtain larger $\Delta_Q$.
Another way we propose for avoiding heating is an adiabatic ramping of $A$ as described in the next section. 

\subsection{\label{sec:adiabatic}Adiabatic ramping of the field amplitude}

\begin{figure*}
\begin{centering}
\includegraphics[width=\hsize]{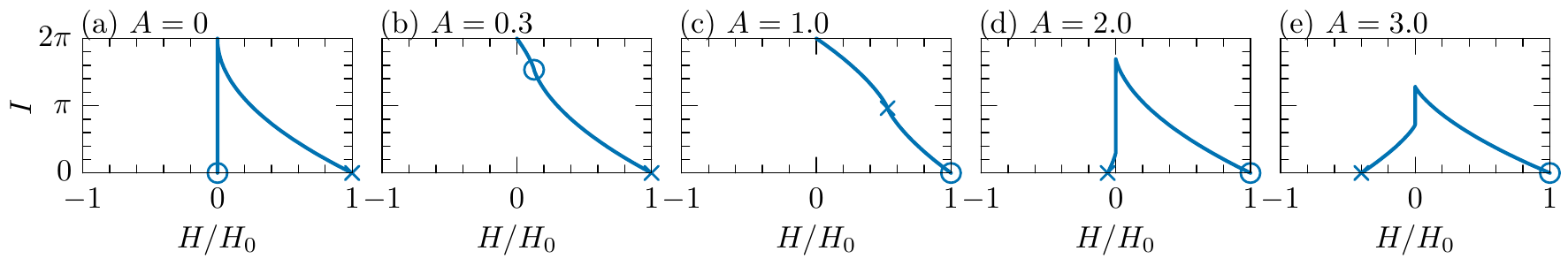}
\caption{\label{fig:adinv}(color online)
The adiabatic invariant $I$ against $H$ [Eq.~(\ref{eq:classical-hamiltonian})] for various values of $A$ employed in Fig.\ref{fig:contour}.
The invariant for the contours that contain $(\rho_Q,\theta)=(0,0)$ ($s$-wave superfluid) are marked with circles,
while those containing $(\rho_Q,\theta)=(0,\pi/2)$ ($\eta$-pairing superfluid) are indicated with crosses. The charge order ($|\rho_Q|=1$) corresponds to $H=0$.
$H$ is normalized here by $H_0=(z/2)(|J_\text{eff}|+|V_\text{eff}|)$.}
\end{centering}
\end{figure*}

Let us alternatively consider an adiabatic change (as opposed to quench) of the amplitude.
We again consider a situation where the initial state is uniform on each sublattice,
where the dynamics is described by the Hamiltonian [Eq.~(\ref{eq:classical-hamiltonian})].
In such a situation we can introduce a classical adiabatic invariant,
\begin{equation}
I=\oint d\theta\rho_{Q}(\theta),
\end{equation}
which is conserved in adiabatic changes of parameters according to the adiabatic theorem~\cite{Landau1977,Liu2002}.
The invariant, measured by the area in the phase space enclosed by the trajectory of a periodic motion,
is conserved as long as the change in the parameter is much slower than the period of the motion. 
While the parameters $J_{ij}(t),V_{ij}(t)$ [Eqs.~(\ref{eq:J-full}), (\ref{eq:V-full})] oscillate rapidly, 
their time averages are approximated by $J_\text{eff}(A(t)),V_\text{eff}(A(t))$.
Namely, $I$ should be conserved approximately in the coarse-grained dynamics as long as the change in the amplitude is slow enough.
Figure~\ref{fig:adinv} illustrates the adiabatic invariant against $H$ for various values of $A$.

We should remember that the conservation of the adiabatic invariant is broken not only for a rapid change of the amplitude 
but even for slow changes when the contour has a saddle point,
at which the period of the motion diverges and the change of the amplitude is no longer adiabatic. 

Now we consider a protocol in which we perform a ramping up of the amplitude slowly,
toward the region where the effective Hamiltonian has an $\eta$-pairing ground state ($-J_{\text{eff}}>V_{\text{eff}}>0$).
Here we take $\omega=0.8U$. As the amplitude is varied,
we walk along a trajectory in the phase diagram displayed by the red curve in Fig.~\ref{fig:phase} aiming at the green region. 
In the initial stage a dynamical instability arises for the $s$-wave superfluid,
which makes the adiabatic invariant $I$ jump from zero to about $2\pi$,
as can be seen in Figs.~\ref{fig:adinv}(a) and \ref{fig:adinv}(b) (marked with circles). 
This can also be seen from the area enclosed by the contour containing $(\rho_Q,\theta)=(0,0)$ ($s$-wave superfluid) in Figs.~\ref{fig:contour}(a) and \ref{fig:contour}(b).
After this, $I$ retains this value during the ramping due to the adiabatic theorem.
This regime corresponds to a charge order 
since the contours [wavy ones along $|\rho_Q|=1$ in Fig.~\ref{fig:contour}~(c)] for this value of $I$ have large $|\rho_Q|$~\footnote{
When the trajectory is open but periodic in $\theta$, $I$ can still be defined as the area between the trajectory and $\rho_Q=0$.}.
The charge order persists as the amplitude is further increased,
until we attain $-J_{\text{eff}}=V_{\text{eff}}$, the phase boundary in Fig.~\ref{fig:phase}.

The situation is then drastically changed due to the charge order collapsing due to the dynamical instability.
We can here interpret this in terms of the adiabatic invariant.
The emergence of the dynamical instability is signaled as a jump in $I$:
As we increase the amplitude, the maximum value of $I$ against $H$ (at $H=+0$) starts to decrease from $2\pi$, as shown in Figs.~\ref{fig:adinv}(d) and \ref{fig:adinv}(e),
so that $I$ can no longer be conserved when the maximum goes below $I$.
When this occurs, the system has $H=+0$, the contour of which is, as seen in Fig.~\ref{fig:contour}(d),
a rectangle (marked with thick lines) enclosing $(\rho_Q,\theta)=(0,0)$.
Its corners exactly correspond to the saddle points (four red circles) triggering the dynamical instability.

On the other hand, there is another side-sharing rectangular contour enclosing the $\eta$-pairing point,
$(\rho_Q,\theta)=(0,\pi/2)$, which belongs to $H=-0$. 
An increase in the amplitude makes the rectangle for $H=+0$ narrower and that for $H=-0$ wider as the shared side (the horizontal thick line) moves away from $\theta=\pi/2$,
as seen in Figs.~\ref{fig:contour}(d) and \ref{fig:contour}(e).
This leads to a decrease of $I(H=+0)$ accompanied by an increase of $I(H=-0)$.

This observation clearly explains the behavior of the system after passing through the phase boundary.
The system is at first in a charge-ordered state with a large $I$. As we increase the amplitude,
the maximum of $I$ decreases until it finally hits the conserved value.
Then the system starts to evolve along the rectangular contour with $H=+0$.
If the amplitude is increased further when the system evolves along the horizontal contour [the thick line in Fig.~\ref{fig:contour}(d)],
the contour drifts away from $\theta=\pi/2$ [see Fig.~\ref{fig:contour}(e)],
and the evolution will be switched to that for $H=-0$ (in the adiabatic limit).
Namely, the adiabatic invariant jumps from $I(H=+0)$ to $I(H=-0)$.
After this, $I$ is conserved with a small value, which results in the large $\eta$-pairing amplitude.

\begin{figure}
\begin{centering}
\includegraphics[width=\hsize]{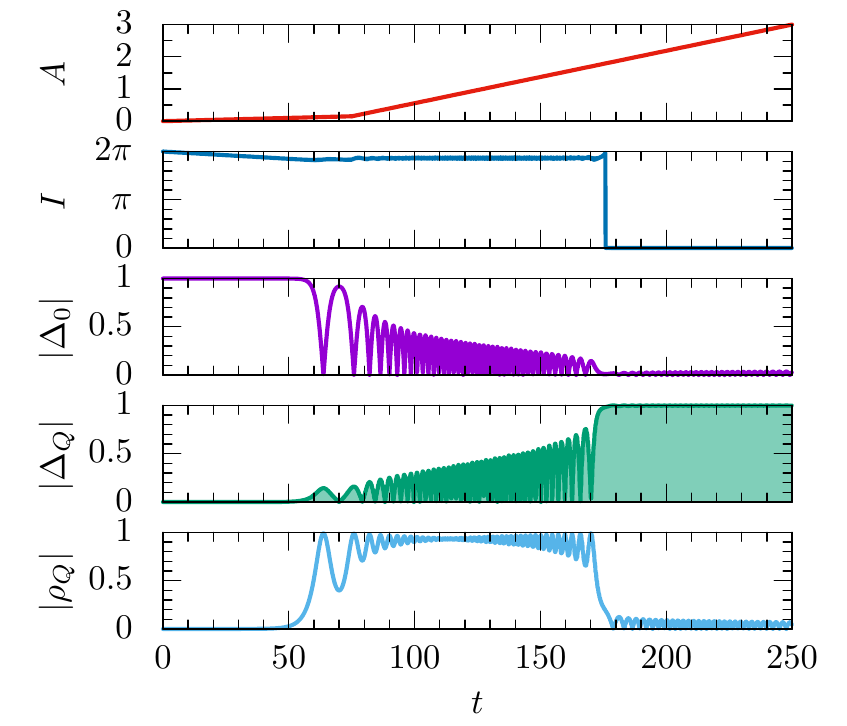}
\caption{\label{fig:adiabatic}(color online) Time evolution of the adiabatic invariant ($I$) and order parameters in an adiabatic protocol for $A$ with $U=10$ and $\omega=8=0.8U$.
The region where $\Delta_Q$ develops is highlighted in light green.
In this simulation, the ramping rate is increased at $t=75$ to shorten the simulation time.
}
\end{centering}
\end{figure}

Let us now turn to the result of a numerical simulation for the actual time evolution in Fig.~\ref{fig:adiabatic}.
We can clearly see that the result indeed excellently agrees with the above argument in that the adiabatic invariant is conserved all the time except at the dynamical instability,
i.e., the situations where saddle points appear and the contours with extremely long periods of motion emerge.
In the present case the charge-ordered state first appears as a result of a large $I$,
and in the second stage a jump of $I$ toward a small value due to the dynamical instability changes the system into an $\eta$-pairing superfluid. 
In this way we end up with a large $|\Delta_Q|$ in this protocol as Fig.~\ref{fig:adiabatic} shows,
which actually provides an alternative method to the two-step quench described in Sec.~\ref{sec:twostep}.

\section{Summary and experimental implications} 
In this paper we have examined dynamical phase transitions in the periodically-driven large $U$ attractive Hubbard model to look for exotic phases.
The effective theory is described by the pair hopping and pair repulsion, which are found to be widely tuned in, both magnitude and sign.
Thanks to a broken $\eta$-SU(2) symmetry induced by the external field,
rich possibilities do emerge, in sharp contrast to the repulsive half-filled case.
We have revealed the emergence of the dynamical instabilities between superfluid phases and charge-ordered phase,
which induce a dynamical phase transition even in isolated systems.
Most importantly, we have invoked dynamical transition as a key process to realize the $\eta$-pairing superfluid,
for which we have proposed and confirmed numerically two protocols,
a two-step quench and an adiabatic ramping of the field intensity.

In this study we have employed several approximations.
One is to consider the leading-order terms in the strong-coupling expansion,
which can be improved by going over to higher orders or by employing the nonequilibrium dynamical mean-field theory~\cite{Aoki2014}.
Another is the mean-field approximation, where we assume that the spatial dimension is sufficiently large.
Applications of the present idea with more quantitative methods or to lower-dimensional systems are interesting future problems.

Let us finally comment on the experimental feasibility,
where most promising candidates are cold-atom systems.
We can manipulate the strength of attractive interaction with the Feshbach resonance,
while the shaken optical lattice has become an established method,
as in an experimental realization of the Floquet topological insulator~\cite{Jotzu2014}. 

Evidence for the $\eta$ pairing proposed in the present paper should be detected in the time-of-flight image of doublons for the cold-atom systems: 
The image directly reflects the distribution of total momenta of the Cooper pairs,
$n_\text{D}(\bm{p})=\langle\hat{\Delta}_{\bm{p}}^\dagger\hat{\Delta}_{\bm{p}}\rangle$, with
$\hat{\Delta}_{\bm{p}}=N^{-1}\sum_i\hat{c}_{i\downarrow}\hat{c}_{i\uparrow}e^{-i\bm{p}\cdot\bm{R}_i}$.
The quantity can be measured in cold-atom experiments,
as has been utilized in detecting fermionic superfluids~\cite{Regal2004}.

\begin{figure}
\begin{centering}
\includegraphics[width=\hsize]{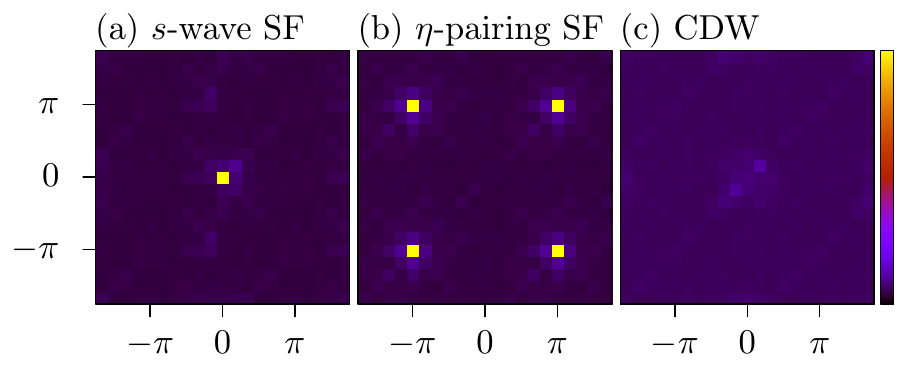}
\caption{\label{fig:tof}(color online) Simulated time-of-flight images for (a) an $s$-wave superfluid, (b) an $\eta$-pairing superfluid, and (c) a charge-density-wave state.
}
\end{centering}
\end{figure}

Figure~\ref{fig:tof} displays simulated time-of-flight images of doublons.
There, we have calculated the total-momentum distribution of doublons (summed over the $z$-axis) for the $s$-wave superfluid, $\eta$-pairing superfluid, and CDW.
To calculate the expectation value of $n_\text{D}(\bm{p})$ we retrieve snapshots of the wave function from the numerical calculation for the amplitude quench (e.g., from $t\sim375$ in Fig.~\ref{fig:biquench}).
We can see in Fig.~\ref{fig:tof} totally distinct features between different phases:
The condensation is signified as sharp peaks at the respective characteristic momenta,
$(0,0)$ for the $s$-wave superfluid and $(\pi,\pi)$ for the $\eta$-pairing superfluid,
while for the charge-ordered phase momenta are uniformly distributed.

\begin{acknowledgments}
The authors wish to thank Philipp Werner and Martin Eckstein for illuminating discussions.
The present work was supported by JSPS KAKENHI Grant No. JP26247057, and by the ImPACT Program of the Council for Science, Technology and Innovation (Cabinet Office, government of Japan; Grant No. 2015-PM12-05-01),
and by the Advanced leading graduate course for photon science (ALPS; S.K.).
\end{acknowledgments}

\appendix*

\section{\label{sec:cano}Derivation of the time-dependent\\ effective Hamiltonian}
Here we derive the time-dependent effective Hamiltonian applicable to nonperiodic amplitude modulations, such as those considered in Sec.~\ref{sec:numerical-time-evolution},
with a time-dependent canonical transformation $e^{-i\hat{S}(t)}$.
Let us denote the hopping term in the original Hamiltonian as $\hat{T}(t)$ and the interaction term as $\hat{V}$ to define $\hat{H}_{\text{D}}(t)$ as 
\begin{align}
\hat{H}_{\text{D}}(t) =\; & e^{i\hat{S}(t)}[\hat{T}(t)+\hat{V}-i\partial_{t}]e^{-i\hat{S}(t)}\nonumber \\
=\; & \hat{V}+\hat{T}(t)+[i\hat{S}(t),\hat{V}]-\partial_{t}\hat{S}(t)+[i\hat{S}(t),\hat{T}]\nonumber \\
& +\frac{1}{2}\left[i\hat{S}(t),[i\hat{S}(t),\hat{V}]-\partial_{t}\hat{S}(t)\right]+\cdots.\label{eq:cano}
\end{align}
In order to determine $\hat{S}(t)$, we regard it as a series in $\hat{T}$ 
and impose a condition that $\hat{H}_{\text{D}}$ has no first-order terms: 
\begin{equation}
\hat{T}(t)+[i\hat{S}(t),\hat{V}]-\partial_{t}\hat{S}(t)=0.\label{eq:cano-1storder}
\end{equation}
This differential equation has a formal solution, 
\begin{equation}
\hat{S}(t)=e^{-it\text{ad}_{\hat{V}}}\hat{S}(0)+\int_{0}^{t}dt^{\prime}e^{-i(t-t^{\prime})\text{ad}_{\hat{V}}}\hat{T}(t^{\prime}),\label{eq:cano-sol}
\end{equation}
where $\text{ad}_{\hat{V}}\bullet=[\hat{V},\bullet]$.
Then we obtain the effective Hamiltonian up to the second order as $\hat{H}_\text{D}(t)=(1/2)\hat{P}[i\hat{S}(t),\hat{T}(t)]\hat{P}$,
where $\hat{P}$ projects out unpaired fermions.
$\hat{S}(0)$ can be determined from a boundary condition $[i\hat{S}(0),\hat{V}]=-\hat{T}(0)$,
where the external field is absent for $t\le0$.
This leads to 
\begin{equation}
\hat{H}_{\text{D}}(t)\sim-\sum_{ij}^{\text{n.n.}}J_{ij}(t)\hat{c}_{i\uparrow}^{\dagger}\hat{c}_{i\downarrow}^{\dagger}\hat{c}_{j\downarrow}\hat{c}_{j\uparrow}+\sum_{ij}^{\text{n.n.}}V_{ij}(t)\frac{\hat{n}_{i}}{2}\frac{\hat{n}_{j}}{2},
\end{equation}
with
\begin{align}
J_{ij}(t)&=\frac{2}{U}e^{i\bm{A}(t)\cdot\bm{R}_{ji}}\cos Ut\notag\\
&+2\int_{0}^{t}dt^{\prime}e^{i[\bm{A}(t)+\bm{A}(t^{\prime})]\cdot\bm{R}_{ji}}\sin U(t-t^{\prime}),\label{eq:J-full}\\
V_{ij}(t)&=\frac{2}{U}\cos[\bm{A}(t)\cdot\bm{R}_{ji}]\cos Ut\notag\\
&+2\int_{0}^{t}dt^{\prime}\cos\{[\bm{A}(t)-\bm{A}(t^{\prime})]\cdot\bm{R}_{ji}\}\sin U(t-t^{\prime}).\label{eq:V-full}
\end{align}

For any boundary condition, the first term in Eq.~(\ref{eq:cano-sol}) results in a contribution like $(1/2)\hat{P}i\hat{S}(0)\hat{T}(t)e^{iUt}\hat{P}$,
which vanishes in the time average [otherwise, the expansion is not justified, as seen in, e.g., the divergence of Eqs.~(\ref{eq:static-JV})],
while the second term leads to $J_\text{eff},V_\text{eff}$ [Eqs.~(\ref{eq:static-JV})] after the time average.
Namely, if the change in amplitude $A(t)$ is sufficiently slow,
the effective Hamiltonian with $J_\text{eff}(A(t)),V_\text{eff}(A(t))$ describes a coarse-grained time evolution which averages out the short timescale $\sim{U^{-1}},\,\omega^{-1}$.

Forms of external fields to which the present formalism is applicable should be restricted from the condition for the expansion to be justified.
For fast components of $\hat{S}(t)$, the Fourier transform of Eq.~(\ref{eq:cano-1storder}) gives $i\hat{S}(\nu){\sim}-\hat{T}(\nu)/\nu$ with a frequency $\nu$,
which should justify the truncation of higher orders in Eq.~(\ref{eq:cano}). 
For slow components, on the other hand, negligible $\partial_{t}\hat{S}(t)$ reduces Eq.~(\ref{eq:cano-1storder}) to that for static problems,
which is justified for strong couplings.
The expansion can break down when the second and third terms in Eq.~(\ref{eq:cano-1storder}) nearly cancel out each other,
unless $\hat{T}(\nu)$ is small enough.
This occurs when $\nu$ is close to $U$, although even in such cases $\hat{T}(\nu)$ is vanishingly small in the small-$A$ region for periodic modulations.

\bibliography{reference}

\end{document}